\documentclass[a4paper,12pt]{article}
\usepackage{epsfig}
\begin{document}
\topmargin 0pt \oddsidemargin 0mm

\begin{titlepage}

\vspace{5mm}
\begin{center}
{\Large \bf Non-commutative Power-law Inflation: Mode Equation and
Spectra Index } \vspace{20mm}

{\large Dao-jun Liu} \vspace{10mm} and
 {\large Xin-zhou Li\footnote{e-mail address:
kychz@shnu.edu.cn}}

\vspace{2mm} {\em  Shanghai United Center for Astrophysics(SUCA),\\
Shanghai Normal University, 100 Guilin Road, Shanghai
200234,China}

\end{center}

\vspace{5mm} \centerline{{\bf{Abstract}}}
 \vspace{5mm}
Following an elegant approach that merge the effects of the
stringy spacetime uncertainty relation into primordial
perturbations suggested by Brandenberger and Ho, we show
 the mode equation up to the first order of non-commutative parameter.
 A new approximation is provided to calculate the mode functions analytically in the
non-commutative power-law inflation models.
 It turns out that non-commutativity of spacetime can provide small
corrections to the power spectrum of primordial fluctuations as
the first-year results of WMAP indicate. Moreover, using the WMAP
data, we obtain the value of expansion parameter, non-commutative
parameter and find the approximation is viable. In addition, we
determined the string scale $l_s \simeq 2.0\times
10^{-29}\mbox{cm}$.

\end{titlepage}

\newpage

The cosmological parameters and the  properties of inflationary
models are tightly constraint by the recent result from Wilkinson
Microwave Anisotropy Probe (WMAP)\cite{WMAP}, Sloan Digital Sky
Survey (SDSS) and Two degree Field (2dF) galaxy clustering
analyses \cite{SDSS}, and from the latest SNIa data \cite{SNIa}.
The standard inflationary $\Lambda$CDM model provides a good fit
to the observed cosmic microwave background (CMB) anisotropies.
The first-year results of WMAP also bring us something intriguing.
Some analyses \cite{Bridle,Mukherjee,Gaztanaga,Kesden} show that
the new data of CMB suggest an anomalously low quadrupole and
octupole and a larger running of the spectral index of the power
spectrum than that predicted by standard single scalar field
inflation models satisfying the slow-roll conditions.

On the other hand, it is well known that during the period of
inflation, the classical gravitational theory, general relativity,
might break down due to the very high energies at that time and
the correction from string theory may take effect. In the
nonperturbative string/M theory, any physical process at the very
short distance takes an uncertainty relation, called stringy
spacetime uncertainty relation (SSUR),
\begin{equation}
\Delta t_p \Delta x_p \geq l_s^2,
\end{equation}
where $t_p$ and $x_p$ are the physical time and space, $l_s$ is
the string length scale. It is suggested that the SSUR is a
universal property for strings as well as D-branes \cite{Yoneya}.
 Unfortunately, we now have no ideas to derive
cosmology directly from string/M theory. Brandenberger and Ho
\cite{Brandenberger} have proposed a variation of spacetime
non-commutative field theory to realize the stringy spacetime
uncertainty relation without breaking any of the global symmetries
of the homogeneous isotropic universe. If the inflation is
affected by physics at a scale close to string scale, one expects
that spacetime uncertainty must leave vestiges in the CMB power
spectrum \cite{Huang,Tsujikawa,Huang2}. It is found that, in the
non-commutative inflation context, IR modes are created on scales
larger than the Hubble radius and thus are not as squeezed as they
would be in the commutative case. Cai~\cite{Cai} show that the
choice of initial vacuum has a significant effect on the power
spectrum of density fluctuation  in a non-commutative spacetime.
Following Ref. \cite{Brandenberger}, the scalar fluctuations of
tachyon inflation was discussed in non-commutative spacetime
\cite{Liu}.

While the standard model is observationally well justified,
successful non-commutative models predict that there should be
observable deviations from it. Undoubtedly, we should expect that
the effect of the non-commutativity of the spacetime may only
provide a small correction to the prediction of the standard
model.

The primary observational test of inflation is observation of CMB.
Temperature fluctuations in the CMB is related to perturbations in
the metric at the surface at last scattering. During the
inflationary epoch, metric perturbations are created by field
fluctuation, and quantum fluctuations on small scales are rapidly
redshifted to scales much larger than the Hubble radius. The
metric perturbations can be decomposed according to their spin
with respect to a local rotation of the spatial coordinates on
hypersurfaces of constant time. This leads to two types: scalar
perturbations which couple to the stress-energy of matter in the
universe and form "seeds" for structure formation, and tensor
perturbations which do not couple to matter.


In this paper, we show the equation of mode functions up to the
first order of non-commutative parameter $\lambda$ beyond the
slow-roll approximation. When the string scale $l_s\rightarrow 0$,
mode equation can be reduced to one in ordinary commutative
spacetime. The mode equation of non-commutative inflation is
complicated, and computing the power spectrum will in general
require numerical evaluation. However, the spectrum can be
evaluated analytically in power-law inflation. There are
corrections to the primordial power spectrum which arise in the
non-commutative power-law models. These corrections lead to a blue
tilt ($n_s>1$) for small wavenumber and a red one ($n_s<1$) for
large wavenumber which accords with the first-year result of WMAP
\cite{WMAP}.

Following the scenario proposed by Brandenberger and Ho
\cite{Brandenberger}, the model incorporating the SSUR can be
written as
\begin{equation}
\label{2eq1}
 S=V\int_{k<k_0}d\tilde \tau d^3k z^2_k(\tilde {\tau})
 (\zeta'_{-k} \zeta'_k-k^2\zeta_{-k}\zeta_k),
 \end{equation}
 where $V$ denotes the total spatial coordinate volume  and the
 primes represent derivatives with respect to the time variable
 $\tilde{\tau}$, which is related to the conformal time $\tau$ via
\begin{equation}\label{tau33}
 d\tilde{\tau}=\left(\frac{a}{a_{eff}}\right)^2d\tau,
\end{equation}
where $a$ is the scale factor, and $a_{eff}$ is defined as
\begin{equation}
a_{eff}\equiv \left(\frac{\beta_k^{+}}{\beta_k^{-}}\right)^{1/4}.
\end{equation}
Here, $\beta_k^{\pm}$ are determined by
\begin{equation}\label{bk}
\beta_k^{\pm}=\frac{1}{2}[a^{\pm2}({\hat{\tau}}+kl_s^2)+a^{\pm2}({\hat{\tau}}-kl_s^2)],
\end{equation}
in which the new time variable $\hat{\tau}$ is defined as
$d\hat{\tau}=a^2d\tau$. $z_k$ in Eq.(\ref{2eq1}) is some smeared
version of the "Mukhanov variable" $z$ over a range of time of
characteristic scale $\Delta \tau=l_s^2 k$,
\begin{equation}
z_k=(\beta_{k}^{+}\beta_k^{-})^{1/4}z=\frac{a\dot{\phi}}{H}(\beta_{k}^{+}\beta_k^{-})^{1/4},
\end{equation}
where $H$ and $\phi$ are Hubble rate and inflaton field,
respectively, and overdot denotes derivative with respect to
cosmic time $t$.

From the action (\ref{2eq1}), the equation of motion of the scalar
perturbations mode equation can be written as
\begin{equation}\label{uk''}
u_k''+\left(k^2-\frac{z_k''}{z_k}\right)u_k=0,
\end{equation}
where the mode function is defined by $u_k=z_k\zeta_k$.

   Apparently, if the string
length scale $l_s$ goes to zero, the action (\ref{2eq1}) will
reduce to the action for the fluctuations in the classical
spacetime, which leads to the equation of motion of perturbations
\begin{equation}
\label{uk''2}
\frac{d^2u_k}{d\tau^2}+\left(k^2-\frac{1}{z}\frac{d^2z}{d\tau^2}\right)u_k=0,
\end{equation}
where $u_k$ now is reduced to $z \zeta_k$. Using the slow-roll
parameters
\begin{equation}
 \epsilon =\frac{M_{pl}^2}{4\pi}\left(\frac{H'(\phi)}{H(\phi)}\right)^2,
\end{equation}
\begin{equation}
 \eta
 =\frac{M_{pl}^2}{4\pi}\frac{H''(\phi)}{H(\phi)},
\end{equation}
\begin{equation}
 \xi
 =\frac{M_{pl}^2}{4\pi}\left(\frac{H'(\phi)H'''(\phi)}{H^2(\phi)}\right)^{1/2},
\end{equation}
the expression for $\frac{1}{z}\frac{d^2z}{d\tau^2}$ can be
written as \cite{Lidsey}
\begin{equation}
\label{ddz}
 \frac{1}{z}\frac{d^2z}{d\tau^2}=2(aH)^2
\left(1+\epsilon-\frac{3}{2}\eta+\epsilon^2-2\epsilon\eta+\frac{1}{2}\eta^2+\frac{1}{2}\xi^2\right).
\end{equation}
The nonlocal coupling in time between the background and the
fluctuation is manifested in Eq.(\ref{bk}). As mentioned above, we
assume that the effect of SSUR only provides a small correction to
the prediction of standard scenario that produce the primordial
fluctuation. This is equivalent to suppose that $k
l_s^2<<|\hat{\tau}|$ in Eq.(\ref{bk}). This condition is crucial
because SSUR takes effects only via $\beta^{\pm}_k$ and it will be
showed that the computations in the following are all based on
this assumption.
 In order to calculate the non-commutative power spectrum
correctly, we introduce a non-commutative parameter $\lambda$ as
\begin{equation}
\lambda(k,t)\equiv\frac{H^2k^2}{a^2M_s^4},
\end{equation}
where $k$ is the comoving wavenumber of a perturbation mode, and
$M_s=l_s^{-1}$ is the string mass scale. There exists a great
difference between the slow-roll parameters $\epsilon$, $\eta$ and
the non-commutative parameter $\lambda$. According to the
definition, slow-roll parameters do not involve $a$ which
increases rapidly during inflation.
Note that $\lambda$ contains scale factor $a$, which is in
contrast to the slow-roll parameters. We note the general picture
of fluctuations during inflation: for a given fluctuation whose
initial wavelength $\sim a/k$ is within the Hubble radius, it
oscillates till the wavelength becomes of the order of the Hubble
radius scale; when the wavelength crosses the Hubble radius, the
fluctuation ceases to oscillate and gets frozen in. After a prolix
but straightforward calculation, we obtain
\begin{equation}
\frac{z_k''}{z_k}=\frac{1}{z}\frac{d^2
z}{d\tau^2}\left[1-2(1+\epsilon)\lambda\right]+2a^2H^2\lambda
\left[3\epsilon+\eta+3\epsilon\eta+\epsilon^2+\epsilon\eta(\epsilon-\eta)\right],\label{ddzk}
\end{equation}
up to the first order of $\lambda$, where
$\frac{1}{z}\frac{d^2z}{d\tau^2}$ is defined in Eq.(\ref{ddz}).
Clearly, when $l_s\rightarrow 0$ or $M_s\rightarrow \infty$, the
quantity $z_k''/z_k$ and $\tilde{\tau}$ will be reduced to
 $\frac{1}{z}\frac{d^2 z}{d\tau^2}$ and $\tau$ respectively, and
 then the mode equation (\ref{uk''}) in non-commutative spacetime will
 recover the one in ordinary commutative spacetime (\ref{uk''2}).

Brandenberger and Ho \cite{Brandenberger} have shown that, for
each mode $k$ of fluctuation, there is a critical time
$\tilde{\tau}_0$ at which the spacetime uncertainty relation is
saturated, and $k$ and $\tilde{\tau}_0$ have the relation
\begin{equation}\label{tau0}
k=\frac{a_{eff}(\tilde{\tau}_0)}{l_s}.
\end{equation}
 $\tilde{\tau}_0$ is also the time when the mode is
generated. Before the critical time $\tilde{\tau}_0$, the
fluctuations do not contain the mode $k$.

Let us now consider power law inflation models  where the scale
factor can be given by $a(\tau)=l_0\mid\tau\mid^{1+\beta}$ where
$\beta$ is a number such that $\beta\leq -2$ and the coefficient
$l_0$ has the dimension of a length. In order for slow-roll
parameter $\epsilon$ to be a little number, we may assume $\beta$
is close to $-2$. In the limit case $\beta=-2$, which corresponds
to exponential expansion, the length $l_0$ is nothing but the
Hubble radius, $l_0=l_H$. Supposing that $l_s<<l_0$, according to
Eq.(\ref{tau0}), we obtain that $-k \tilde{\tau}_0 \approx
{l_0}/{l_s}>>1 $ provided that $\beta$ is close to $-2$, which
means that the mode $k$ is generated on scales inside the Hubble
radius in the local vacuum state.

In power law models, the slow-roll parameters can all be
determined exactly,
\begin{equation}
\epsilon=\eta=\xi=\frac{2+\beta}{1+\beta},
\end{equation}
 which is a virtue of this class of models, and
the mode equation (\ref{uk''}) thus is reduced to
\begin{equation}\label{uk''3}
\frac{d^2u_k}{d\tilde{\tau}^2}+\left[k^2-\beta(1+\beta)\frac{1}{\tau^2}
+{4\alpha k^2}
(1+\beta)^2\left(5+5\beta+\beta^2\right)\frac{1}{|\tau|^{8+4\beta}}\right]u_k=0,
\end{equation}
where the parameter $\alpha\equiv\left(\frac{l_s}{l_0}\right)^4$,
and the relation between conformal time $\tau$ and $\tilde{\tau}$
can be rewritten as
\begin{equation}\label{tau`}
\tilde{\tau}=\tau+\alpha
k^2\frac{(1+\beta)(3+2\beta)}{(5+4\beta)}\frac{1}{|\tau|^{5+4\beta}}.
\end{equation}

 In principle, we can solve Eq.(\ref{tau`}), then
insert the solution $\tau(\tilde{\tau})$ into Eq.(\ref{uk''3}) and
finally obtain the solution $u_k(\tilde{\tau})$ of
Eq.(\ref{uk''3}). However, this procedure is too complicated to
implement directly in practice. The situation here are in many
ways equivalent to the modified dispersion relations considered by
Brandenberger and Martin in ref.\cite{Martin}. Note that
Eq.(\ref{uk''3}) is a linear equation and we have assumed $\alpha
<<1$. Therefore, we can use perturbation method to solve the
Eqs.(\ref{uk''3}) and (\ref{tau`}). For this purpose, let
\begin{equation}\label{tau}
\tau=\tau^{(0)}+\alpha\tau^{(1)}+\alpha^2\tau^{(2)}+\cdots,
\end{equation}
\begin{equation}\label{u_k}
u_k=u_k^{(0)}+\alpha u_k^{(1)}+\alpha^2 u_k^{(2)}+\cdots.
\end{equation}
Inserting them into Eqs.(\ref{uk''3}) and (\ref{tau`}), we obtain
that
\begin{equation}
\tau=\tilde{\tau}-\alpha
k^2\frac{(1+\beta)(3+2\beta)}{(5+4\beta)}\frac{1}{|\tilde{\tau}|^{5+4\beta}}+O(\alpha^2),
\end{equation}
and then

\begin{equation}\label{uk0''}
u_k''^{(0)}+\left[k^2-\frac{\beta(1+\beta)}{\tilde{\tau}^2}\right]u_k^{(0)}=0,
\end{equation}
\begin{equation}\label{uk1''}
u_k''^{(1)}+\left[k^2-\frac{\beta(1+\beta)}{\tilde{\tau}^2}\right]u_k^{(1)}
=h(\tilde{\tau}),
\end{equation}
where
\begin{equation}
h(\tilde{\tau})=\frac{2k^2(1+\beta)^2}{|\tilde{\tau}|^{8+4\beta}}\left[\frac{\beta(3+2\beta)}{(5+4\beta)}-2\left(5+5\beta+\beta^2\right)\right]u_k^{(0)}.
\end{equation}
 If the non-commutative spacetime effects are ignored, the
mode function $u_k$ is reduced to $u_k^{(0)}$ which obeys
Eq.(\ref{uk0''}). It is easy to find that Eq.(\ref{uk0''}) is
nothing but Eq.(\ref{uk''2}) with $\tau$ replaced by
$\tilde{\tau}$. Thus, if we impose that in the ultraviolet regime
($-\tilde{\tau}_0>-\tilde{\tau}>>1/k$), the solution of
Eq.(\ref{uk0''}) matches the plane-wave solution we expect in flat
spacetime and obey the Wronskian condition
\begin{equation}
u_k^{*}\frac{du_k}{d\tilde{\tau}}-u_k\frac{du_k^{*}}{d\tilde{\tau}}=-i,
\end{equation}
 the exact solution of Eq.(\ref{uk0''})
becomes
\begin{equation}\label{uk(0)}
u_k(\tilde{\tau})^{(0)}=\frac{\sqrt{\pi}}{2}\exp\left[i\frac{\pi}{2}(\nu+\frac{1}{2})\right](-\tilde{\tau})^{1/2}H^{(1)}_{\nu}(-k\tilde{\tau}),
\end{equation}
where, the parameter $\nu=-\frac{1}{2}-\beta$. For solving
Eq.(\ref{uk1''}), we can use the general methods of both
homogeneous and inhomogeneous linear ordinary differential
equations. The general solutions of the second-order equation can
be written as
\begin{equation}\label{uk(1)}
u_k(\tilde{\tau})^{(1)}=\varphi_2\int\frac{\varphi_1h}{W}d\tilde{\tau}-\varphi_1\int\frac{\varphi_2h}{W}d\tilde{\tau},
\end{equation}
where two linearly independent solutions of the homogeneous
equation
\begin{equation}
\varphi_1=(-\tilde{\tau})^{1/2}H^{(1)}_{\nu}(-k\tilde{\tau}),
\end{equation}
\begin{equation}
\varphi_2=(-\tilde{\tau})^{1/2}H^{(2)}_{\nu}(-k\tilde{\tau}),
\end{equation}
and the Wronskian of the two solutions of homogeneous equation
\begin{equation}
W=\varphi_2\frac{d\varphi_1}{d\tilde{\tau}}-\varphi_1\frac{d\varphi_2}{d\tilde{\tau}},
\end{equation}
Fortunately, the integral in Eq.(\ref{uk(1)}) can be explicitly
integrated.

On the subhorizon scales, i.e. for
$k^2\tilde{\tau}_0^2>>k^2\tilde{\tau}^2>>1$, since $\varphi_1\sim
\sqrt{\frac{2}{\pi
k}}e^{-i(k\tilde{\tau}+\frac{\pi}{2}\nu+\frac{\pi}{4})}$,
$\varphi_2\sim \sqrt{\frac{2}{\pi
k}}e^{i(k\tilde{\tau}+\frac{\pi}{2}\nu+\frac{\pi}{4})}$ and
$u_k^{(0)}\sim \frac{1}{\sqrt{2k}}e^{-ik\tilde{\tau}}$, inserting
these expressions into Eq.(\ref{uk(1)}), we obtain that
\begin{equation}
u_k^{(1)}\approx ik
\frac{(1+\beta)^2(50+87\beta+48\beta^2+8\beta^3)}{(5+4\beta)(7+4\beta)}(-\tilde{\tau})^{-7-4\beta}
\frac{e^{-ik\tilde{\tau}}}{\sqrt{2k}}.
\end{equation}
We are specially interested in solution on the superhorizon
scales, i.e. for $k^2\tilde{\tau}^2<<1$. For these scales, since
$H_{\nu}^{(1)}(x<<1)=\sqrt{\frac{2}{\pi}}e^{-i\pi/2}2^{\nu-3/2}\frac{\Gamma(\nu)}{\Gamma(3/2)}x^{-\nu}$,
we obtain that
\begin{equation}
u_k^{(0)}\approx
e^{i(\nu-\frac{1}{2})\frac{\pi}{2}}2^{\nu-\frac{3}{2}}\frac{\Gamma(\nu)}{\Gamma(3/2)}\frac{1}{\sqrt{2k}}(-k\tilde{\tau})^{\frac{1}{2}-\nu},
\end{equation}
and
\begin{equation}
u_k^{(1)}\approx -\frac{{{(1+\beta )}^2} (50 + 87\beta + 48\beta^2
+ 8\beta^3)}{{\sqrt{\pi }}(75 + 140\beta + 84\beta^2 + 16\beta^3)}
{2^{-\frac{3}{2}-\beta }} {{e }^{-i\frac{\pi}{2}( \beta+1)
}}\Gamma\left(-{1}/{2}-\beta\right)  {{ k}^{\frac{5}{2}+\beta }}
{(-\tilde{\tau})^{-5-3 \beta }} .
\end{equation}

Therefore, we can express the power spectrum on superhorizon
scales of the comoving curvature as
\begin{eqnarray}
P_R(k)&=&\frac{k^3}{2\pi^2}\left|\frac{u_k(\tilde{\tau}_c)}{z_k(\tilde{\tau}_c)}\right|^2
\nonumber\\
&\simeq&\frac{k^3}{2\pi^2}\frac{u_k^{(0)}u_k^{(0)*}+\alpha \left(
u_k^{(0)}u_k^{(1)*}+u_k^{(0)*}u_k^{(1)}\right)}{z_k^2}\Bigg|_{\tilde{\tau}=\tilde{\tau}_c}
\nonumber\\
&\simeq&\frac{2^{-2\beta-2}{\Gamma(\nu)}^2k^{4+2\beta}}{{\pi^2}M_{pl}^2l_0^2\epsilon}
\left[1+2\alpha f(\beta) k^2(-\tilde{\tau}_c)^{-6-4\beta}\right],
\end{eqnarray}
where
\begin{equation}
f(\beta)=-(1+\beta)^2\left(1+\frac{50+87\beta+48\beta^2+8\beta^3}{75+140\beta+84\beta^2+16\beta^3}\right),
\end{equation}
and $\tilde{\tau}_c$ is the time when fluctuation mode $k$ comes
across the Hubble radius, (i.e. for $-\tilde{\tau}_c \approx
1/k$). Just as Lidsey \textit{et al} have pointed in
Ref.\cite{Lidsey} that, in spite of the appearance of spectrum
equation, the calculated value for the spectrum is not the value
at which the scale crosses outside the Hubble radius. Rather, it
is the asymptotic value as $k/aH\rightarrow 0$, but rewritten in
terms of the values the quantities had when the Hubble radius was
crossed.

 We may now compute the spectra index $n_s$ of the scalar
metric perturbation on superhorizon scales
\begin{equation}
n_s-1\equiv \frac{d\ln P_R(k)}{d\ln
k}\approx{2(2+\beta)\left[1+4\alpha f(\beta)k^{8+4\beta}\right]}.
\end{equation}
The running of the spectrum index is
\begin{equation}
\frac{dn_s}{d\ln
k}\approx{32\alpha(2+\beta)^2f(\beta)k^{8+4\beta}}.
\end{equation}
Obviously, when the parameter $\alpha\rightarrow 0$ (i.e.
$M_s\rightarrow \infty$), the contribution from the
non-commutativity of spacetime to the spectral index and its
running will also vanish. Note that in the vicinity of $\beta=-2$,
$f(\beta)$ is negative. Thus, the spectrum has a negative spectral
index for small scales and a positive one for large scales (see
Figure 1), while the running is always negative. Since the slope
of the power spectrum decreases as $\beta$ goes towards to $-2$,
the more rapidly the universe is accelerating, the closer the
power spectrum is to being scale-invariant. In the limit case
$\beta=-2$, the results for commutative and non-commutative
spacetimes converge at a completely scale-invariant spectrum.

\begin{figure}
\epsfig{file=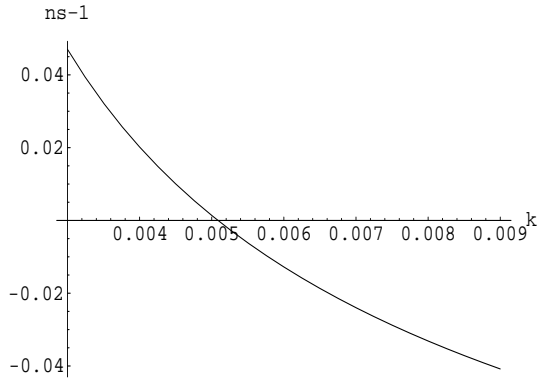,height=2.0in,width=2.8in} \caption{The
spectral index $n_s-1$ for different wavenumber $k$, where the
parameter $\alpha=10^{-2}$ and $\beta=-2.1$.}
\end{figure}

In the non-commutative  inflationary spacetime, there are
corrections to the primordial power spectrum which arise in a
model of power-law inflation. These corrections lead to a blue
tilt ($n_s>1$) for small wavenumber and a red one ($n_s<1$) for
large wavenumber which accords with the first-year results of
WMAP\cite{WMAP}. The origin of the suppressions in the power
spectrum of the fluctuations is that the noncommutativity of the
spacetime delayed the generation of the fluctuation modes and then
postponed the time when they crossing the Hubble radius. However,
in the de Sitter limit i.e. for $\epsilon=\eta=\xi=0$, the
non-commutativity of the spacetime has no influence in the
spectrum, this is because no time delay can be generated in this
special case. According to the analysis of the results of
WMAP\cite{WMAP,Huang}, for the scalar modes, the mean and the
$68\%$ error level of the 1-d marginalized likelihood for the
power spectrum slope $n_s=0.93^{+0.02}_{-0.03}$, $dn_s/d\ln
k=-0.031^{+0.016}_{-0.017}$ at $k=0.05 \mbox{Mpc}^{-1}$ and
$n_s=1.20^{+0.12}_{-0.11}$, $dn_s/d\ln k=-0.077^{+0.050}_{-0.052}$
at $k=0.002\mbox{Mpc}^{-1}$. Using the data at $k=0.05
\mbox{Mpc}^{-1}$, the parameters $\beta$ and $\alpha$ should be
constraint by $\beta\simeq -2.08$ and $\alpha\simeq 0.0186$,
respectively. The parameter $\alpha$ is so small that ensure that
our treatments , i.e. Eqs. (\ref{tau}) and (\ref{u_k}), are
suitable. Using the values of parameters $\beta$ and $\alpha$
gained above, we predict that $n_s\simeq1.11$, $dn_s/d\ln
k\simeq-0.089$ at the scale of $k=0.002\mbox{Mpc}^{-1}$. This
results are in good agreement with those obtained in
Ref.\cite{Huang}.  Although the predicted central values of the
spectra index and its running have small deviations from the
corresponding WMAP data, but they both fall within the error bar.
The differences exist due to the fact that at the large scales,
the effect of higher orders in $\alpha$ is not completely
negligible. As these higher-order effects is taken into account,
the result would be improved. In addition, using the WMAP data
that $P_R(k=0.002\mbox{Mpc}^{-1})=2.09\times10^{-9}$ and the
parameters obtained above, we estimate the string scale $l_s
\simeq 1.2\times 10^4 l_{p}\simeq 2.0\times 10^{-29}\mbox{cm}$,
which is also consistent with the result obtained in
Ref.\cite{Tsujikawa,Huang2}.

In summary, following the elegant idea that merge the effects of
the stringy spacetime uncertainty relation into primordial
perturbations proposed by Brandenberger and Ho, we obtain
 the mode equation up to the first order of non-commutative
 parameter. Moreover, we also provide a new analytical approximation to calculate the mode functions in the
 power-law inflation models. It turns out that non-commutativity of spacetime can provide small
corrections to the power spectrum of primordial fluctuations and
our results is consistent with the previous results and WMAP data.

\section*{Acknowledgments} This work was partially supported by
NKBRSF under Grant No. 1999075406.


\begin{thebibliography}{99}
\bibitem{WMAP}C.~L.~Bennett {\it et al.},
Astrophys.\ J.\ Suppl.\  {\bf 148}, (2003) 1 ; D.~N.~Spergel {\it
et al.},
 Astrophys.\ J.\ Suppl.\  {\bf 148}, (2003) 175
; G.~Hinshaw {\it et al.},
 Astrophys.\ J.\ Suppl.\  {\bf 148}, (2003) 135.


\bibitem {SDSS} M. Tegmark {\it et al.}, Astrophys. J. \textbf{606} (2004)
702.
\bibitem {SNIa} A. G. Riess {\it et al.},Astrophys. J. \textbf{607}
(2004) 665.
\bibitem{Bridle}S. L. Bridle, A. M. Lewis, J. Weller and G.
Efstathiou, MNRAS, \textbf{342}, (2003) L72 .
\bibitem{Mukherjee}P. Mukherjee, Y. Wang, Astrophys. J. \textbf{599},(2003) 1
.
\bibitem{Gaztanaga} E. Gaztanaga, J. Wagg, T. Multamaki, A. Montana and D. H.
Hughes, MNRAS, \textbf{346}, (2003) 47 .

\bibitem{Kesden} M. Kesden, M. Kamionkowski and A. Cooray, Phys. Rev. Lett. \textbf{91},
(2003) 221302 ; M. Kesden, A. Cooray and M. Kamionkowski, Phys.
Rev. \textbf{D67}, (2003) 123507.

\bibitem {Yoneya} T. Yoneya, in "Wandering in the Fields", eds. K.
Kawarabayashi, A. Ukawa (World Scientific, 1987), P. 419; M. Li
and T. Yoneya, Phys. Rev. Lett. \textbf{78}, (1997) 1219 ; T.
Yoneya, Prog. Theor. Phys. 103, (2000) 1081 .

\bibitem {Brandenberger} R. Brandenberger and P. M. Ho, Phys. Rev.
\textbf{D66}, (2002) 023517 .

\bibitem {Huang} Q. G. Huang and M. Li, JHEP \textbf{0306}, (2003) 014 .
\bibitem {Tsujikawa} S. Tsujikawa, R. Maartens and R.
Brandenberger, Phys. Lett. \textbf{B574}, (2003) 141 .
\bibitem {Huang2} Q. G. Huang and M. Li, JCAP \textbf{0311}, (2003) 001.

\bibitem {Cai} R. G. Cai, Phys. Lett. \textbf{B593} (2004) 1.

\bibitem {Liu} D. J. Liu and X. Z. Li, astro-ph/0402063;
\\X. Z. Li,
J. G. Hao and D. J. Liu, Chin. Phys. Lett. \textbf{19} (2002)
1584, hep-th/0204252.

\bibitem {Lidsey}J. E. Lidsey , A. R. Liddle , E. W. Kolb , E. J. Copeland , T. Barreiro  and M. Abney, Rev. Mod. Phys. \textbf{69}
(1997) 373 .
\bibitem {Martin} J. Martin and R. H. Brandenberger, Phys. Rev.
\textbf{D63}, (2001) 123501 .

\end{thebibliography}
\end{document}